\documentclass[reprint,superscriptaddress,
prb,
amsmath,amssymb,longbibliography]{revtex4-2}
\usepackage{graphicx}
\usepackage{dcolumn}
\usepackage{sidecap}
\usepackage{bm}
\usepackage{color}
\usepackage{textgreek}
\usepackage{braket}
\usepackage{xcolor} 

\usepackage{hyperref}
\usepackage{amsmath}
\usepackage{titlesec} 
\titlespacing\section{0pt}{10pt}{4pt}
\titlespacing\subsection{0pt}{10pt}{2pt}
\usepackage{soul}

\begin{document}
\title{Near zero-field microwave-free magnetometry with ensembles of nitrogen-vacancy centers in diamond}

\author{Omkar Dhungel}
\affiliation{Helmholtz-Institut Mainz, GSI Helmholtzzentrum f{\"u}r Schwerionenforschung GmbH, 55128 Mainz, Germany}
\affiliation{Johannes Gutenberg-Universit{\"a}t Mainz, 55128 Mainz, Germany}

\author{Till Lenz}
\affiliation{Helmholtz-Institut Mainz, GSI Helmholtzzentrum f{\"u}r Schwerionenforschung GmbH, 55128 Mainz, Germany}
\affiliation{Johannes Gutenberg-Universit{\"a}t Mainz, 55128 Mainz, Germany}

\author{Muhib Omar}
\affiliation{Helmholtz-Institut Mainz, GSI Helmholtzzentrum f{\"u}r Schwerionenforschung GmbH, 55128 Mainz, Germany}
\affiliation{Johannes Gutenberg-Universit{\"a}t Mainz, 55128 Mainz, Germany}

\author{Joseph Shaji Rebeirro}
\affiliation{Helmholtz-Institut Mainz, GSI Helmholtzzentrum f{\"u}r Schwerionenforschung GmbH, 55128 Mainz, Germany}
\affiliation{Johannes Gutenberg-Universit{\"a}t Mainz, 55128 Mainz, Germany}

\author{Minh-Tuan Luu}
\affiliation{Max-Planck Institut f{\"u}r Polymerforschung, Ackermannweg 10, 55128 Mainz, Germany}

\author{Ali Tayefeh Younesi}
\affiliation{Max-Planck Institut f{\"u}r Polymerforschung, Ackermannweg 10, 55128 Mainz, Germany}

\author{Ronald Ulbricht}
\affiliation{Max-Planck Institut f{\"u}r Polymerforschung, Ackermannweg 10, 55128 Mainz, Germany}

\author{Viktor Iv{\'a}dy}
\affiliation{\it Department of Physics of Complex Systems, ELTE E{\"o}tv{\"o}s Lor{\'a}nd University, Egyetem t{\'e}r 1-3, H-1053 Budapest, Hungary}
\affiliation{MTA-ELTE Lend{\"u}let “Momentum” NewQubit Research Group, P{\'a}zm{\'a}ny P{\'e}ter, S{\'e}t{\'a}ny 1/A, 1117 Budapest, Hungary}
\affiliation{Department of Physics, Chemistry and Biology, Link{\"o}ping University, 581 83 Link{\"o}ping, Sweden}

\author{Adam Gali}
\affiliation{Wigner Research Centre for Physics, P.O.\ Box 49, H-1525 Budapest, Hungary}
\affiliation{Budapest University of Technology and Economics, Institute of Physics, Department of Atomic Physics, M\"{u}egyetem rakpart 3., 1111 Budapest, Hungary}

\author{Arne Wickenbrock}
\affiliation{Helmholtz-Institut Mainz, GSI Helmholtzzentrum f{\"u}r Schwerionenforschung GmbH, 55128 Mainz, Germany}
\affiliation{Johannes Gutenberg-Universit{\"a}t Mainz, 55128 Mainz, Germany}

\author{Dmitry Budker}\email{budker@uni-mainz.de}
\affiliation{Helmholtz-Institut Mainz, GSI Helmholtzzentrum f{\"u}r Schwerionenforschung GmbH, 55128 Mainz, Germany}
\affiliation{Johannes Gutenberg-Universit{\"a}t Mainz, 55128 Mainz, Germany}
\affiliation{Department of Physics, University of California, Berkeley, California 94720-300, USA}

\date{\today}

\begin{abstract}  
\setstcolor{red}

We study cross-relaxation features near zero magnetic field with ensembles of nitrogen-vacancy (NV) centers in diamond and examine their properties in samples with a range ($0.9$\,ppm\,-\,$16.0$\,ppm) of NV concentrations. The observed NV-NV cross-relaxation features between differently oriented NV centers in high ($\gtrsim0.9$\,ppm)-NV-density samples hold promise for a variety of magnetometry applications where microwave fields (or any bias field) disturb the system under study. We theoretically determine the values of the bias magnetic fields corresponding to cross-relaxations between different axes and experimentally validate them. The behavior of zero-field cross-relaxation features as a function of temperature is also investigated.

\end{abstract}

\maketitle 

\section{Introduction}
\setstcolor{red}

Nitrogen-vacancy (NV) centers in diamond are used in applications such as magnetic field\,\cite{Rondin2014}, electric field \,\cite{dolde2011electric} and temperature sensors\,\cite{doi:10.1021/nl401216y} over a wide range of environmental conditions, e.g., pressure and temperature\,\cite{CHOE20181066, PhysRevLett.112.047601}. In most applications as magnetometers, the magnetic field is determined by using optically detected magnetic resonance (ODMR)\,\cite{PhysRevB.42.8605}. In this measurement protocol, the diamond is illuminated with green laser light and a microwave field of tunable frequency is applied to measure the energy separation between the magnetically-sensitive spin 1 ground-state levels to obtain the magnetic field value. For measuring weak magnetic fields where Zeeman shifts of sublevels may be nonlinear in field\,\cite{Rondin2014}, this method usually requires a bias field to lift the degeneracy between the $m_s=\pm1$ sublevels.
However, the use of microwave and external bias magnetic fields may disrupt the system of interest in some applications. To overcome this limitation, different strategies emerged. Recently, microwave-free magnetometry\,\cite{doi:10.1063/1.4960171, https://doi.org/10.1002/qute.202000037, 10.1063/5.0059330},  as well as vector magnetometry\,\cite{PhysRevApplied.13.044023} based on the level anti-crossing in the triplet ground state at 102.4\,mT have been demonstrated. Zero-field magnetometry was realized for both ensemble and single NV centers by using circularly polarized microwave fields to individually address transitions to the $m_s=+1$ or $m_s=-1$ states \,\cite{PhysRevApplied.11.064068,lenz2021}. Still, due to the high external magnetic field or the application of microwave fields, these techniques may be problematic for studying systems where both external magnetic field and the microwaves might disturb the system. Examples of such systems are high-T$_c$ superconductors (T$_c$ stands for the superconducting transition temperature)\,\cite{PhysRevApplied.10.034032}, zero- to ultra-low field NMR\,\cite{BLANCHARD2021106886}, biological samples and various magnetic materials\cite{PhysRevApplied.15.024040, simpson2016magneto}. 

 Recently, ODMR experiments with an additional radiofrequency field were used to study features occurring in fluorescence measurements at zero and low fields \,\cite{Holliday1989,dmitriev2019high,PhysRevA.105.043509}. In addition, microwave-free magnetometry at low field was proposed. The technique probes the position of the cross-relaxation resonances\,\cite{PhysRevA.94.021401, Anishchik2015, PhysRevB.40.6509} and allows for vector magnetometry\,\cite{PhysRevA.96.013806, PhysRevA.100.043844}. This new magnetometry technique therefore overcomes the limitations of microwave-based protocols.

Motivated by the desire to improve the sensitivity of near-zero-field magnetometry,  in this work, we perform a detailed study of the cross-relaxation features with respect to the parameters that may affect the contrast and width of the feature. These include the orientation of the magnetic field with respect to the crystal axes, NV density (Sec.\,\ref{characterization}), and the sample temperature (Sec.\,\ref{temperature}). We found that a boost in sensitivity can be achieved by applying a transverse bias field, therefore, we study the cross-relaxations under transverse fields up to $\approx2.0$\,mT as a function of the azimuthal angle (Sec.\,\ref{transverse_field}). We numerically predict and experimentally verify the observed cross-relaxation patterns. Conclusions and outlook are given in Sec.\,\ref{conclusion}.

\section{NV-center ground state}
\label{Sec:NV_GS}

\begin{figure}

\includegraphics[width=3.0in]{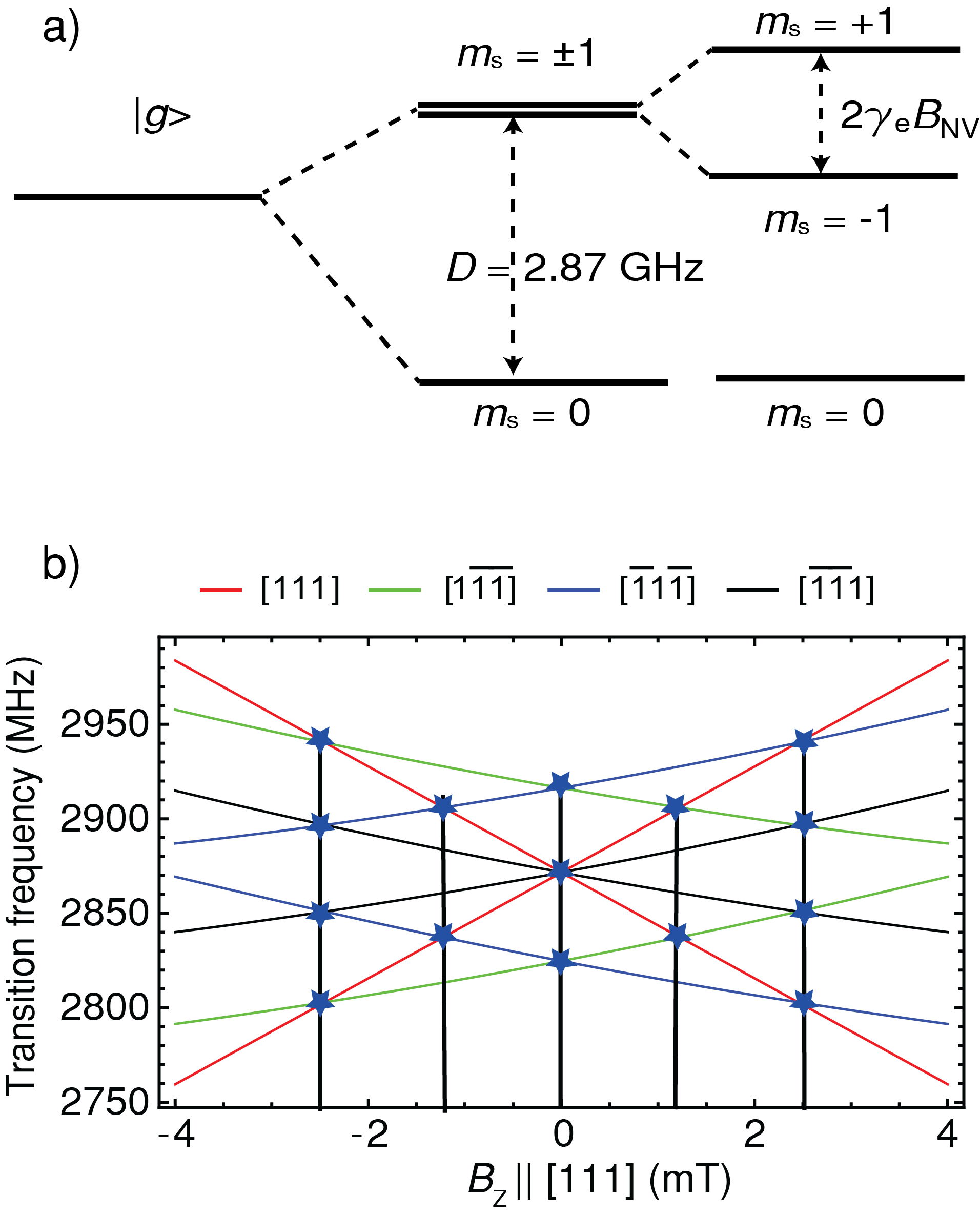}
\caption{(a) Energy-level diagram of the NV center in the presence of magnetic field. (b) Energy separation (transition energy) between the  ground $m_s=0$ and $\pm 1$ states for all four crystallographic axes as a function of $B_z$ when a 2\,mT transverse field is applied along $\hat{x}$. Blue stars are the positions where the crossings happen between transition energies for different NV orientations. Fifteen transition-energy crossing between different orientations are expected in total.}
 \label{energy level}
\end{figure}

The fluorescence rate is reduced due to the ${T_1}$ relaxation between bright and dark states. The coupling mediated by dipolar coupling leads to faster ${T_1}$ relaxation and such coupling is enhanced when NV states of different axes are degenerate\,\cite{Anishchik2015}. In the following, we are studying the Hamiltonian to determine at what external fields this might be observable and compare it to the experiment. The NV ground state in the presence of an external magnetic field can be described by the Hamiltonian: 
\begin{equation}\label{eq1}
H = D(S_z^2- \frac{1}{3}\vec{S}^2) + E(S_x^2-S_y^2) +g_s\mu_{B}\vec{B} \cdot \vec{S}\,,
\end{equation}
where $D$= 2.87\,GHz is the  axial and $E$ is the transverse zero-field-splitting (ZFS) parameters, $\vec{B}$ is the magnetic field, $\vec{S}$ is the electronic spin with components $S_{x}$, $S_{y}$ and $S_{z}$. The electron $g$ factor is $g_s$ =\,2.003 \cite{PhysRevB.79.075203} and $\mu_{B}$ is the Bohr magneton. The Larmor frequency of the  NV center is $\Omega$ = $\frac{g_s\mu_{B}}{{\hbar}}\,B$. Here, the  $z$-axis is chosen along the NV axis. The spin operators $S_{x}$, $S_{y}$, $S_{z}$ can be written as:
\begin{equation}
\begin{pmatrix}
0 & \frac{1}{{\sqrt{2}}} & 0\\
\frac{1}{{\sqrt{2}}} & 0 & \frac{1}{{\sqrt{2}}} \\
0 & \frac{1}{{\sqrt{2}}} & 0
\end{pmatrix},
\begin{pmatrix}
0 & \frac{-i}{\sqrt{2}} & 0\\
\frac{i}{{\sqrt{2}}} & 0 & \frac{-i}{{\sqrt{2}}} \\
0 & \frac{1}{{\sqrt{2}}} & 0
\end{pmatrix},  \begin{pmatrix}
1 & 0 & 0\\
0 & 0 & 0 \\
0 &0 & -1
\end{pmatrix}\,.
\end{equation}
A unit vector with an angle of $\beta$ to the $z$ axis can be written as:
\begin{equation} \label{}
\vec{n} =(\cos\phi \sin\beta, \sin\beta \sin\phi , \cos\beta),
\end{equation} 
where $\phi$ is the azimuthal angle.
Then eq.\,\eqref{eq1} can be written as:
\begin{equation} \label{eq2}
    H= \begin{pmatrix}
D+ \Omega \cos\beta & \frac{ e^{-i\phi}\Omega \sin\beta}{{\sqrt{2}}} & E\\
\frac{e^{i\phi}\Omega \sin\beta}{\sqrt{2}} & 0 & \frac{e^{-i\phi}\Omega \sin\beta}{{\sqrt{2}}} \\
E & \frac{e^{i\phi}\Omega \sin\beta}{\sqrt{2}} & D- \Omega \cos\beta\,
\end{pmatrix}.
\end{equation}
A schematic of the energy levels obtained by diagonalization of the Hamiltonian \eqref{eq2} is shown in Fig.\,\ref{energy level}. The ground state $\ket {g}$ is split by the dipole-dipole interaction described by the parameter $D$, lifting the degeneracy between the $m_s$= $\pm$1 and $m_s$= 0 sublevels. Application of a magnetic field along NV-axis further lifts the $m_s= \pm 1$ degeneracy level by 2$\gamma_eB_{\textrm{NV}}$ above the $m_s$= -1 when $B_{\textrm{NV}}$ is applied.

\begin{table*}[htpb]
    \centering
    \begin{tabular*}{\linewidth}{@{\extracolsep{\fill}} lcccc cccl}
    \hline 
    \hline

    Diamond Name & Cut & [N] density\,(ppm) & NV density\,(ppm) & Laser intensity\,(mW) & $\gamma$\,(mT) & Optimum C\,(\%) & $\gamma$\,(mT)/ C\,(\%)\\
    \hline

    Super diamond & 111 & $\sim$3 & 0.9 & 5 & 0.12 & 0.04 & 3.00\\

    George & 100 & $\sim$13 & 3.7 & 5.93 & 0.20 & 0.91 & 0.21\\
    
    1970608-29 & 111 & - & 3.8 & 2.98 & 0.78 & 2.85 & 0.27\\
    
    S2 & 111 & $<$100 & 16.0 & 20.2 & 1.42 & 4.01 & 0.35\\
    
    sumi\_300 kev* & 110 & $\sim$150 & 14.0 & 37.00 & 1.42 & 1.21 & 1.17\\

    2170612-13*  & 100 & $\sim$14 & 4.5 & 37.00 & 0.17 & 0.29 & 0.59\\
    \hline 
    \hline
    
    \end{tabular*}\
    \caption{Characterization of the zero-field cross-relaxation feature in different diamond samples where $B_z$ is along the cut surface. Here, $\gamma$ and C are the linewidth and contrast, respectively. The samples with * in their names are thin-layer samples.}
    \label{bulk}
\end{table*}

Consider the following four possible crystallographic axes of diamond lattice for NV centers: [111], [1$\overline{1}\overline{1}$], [$\overline{1}1\overline{1}$] and [$\overline{1}\overline{1}$1]. In the diamond samples used in this work, the NV centers are uniformly distributed over the possible orientations. Since NVs of all orientations have equal ground state splitting, the transition energies of all orientations cross at $\vec{B}=0$. When the splittings of orientationally inequivalent NV centers match, a cross-relaxation feature occurs and we observe a decrease in fluorescence. A recently developed model\,\cite{PhysRevLett.118.093601} suggests that local energy relaxation occurs when a randomly distributed portion of NV centers rapidly incoherently depolarizes. Through dipolar interactions, these spins can depolarize the entire ensemble at zero-field. It was also observed that there is a significant contribution of local electric fields and the interaction between same and differently oriented NVs on depolarization at zero-field for high-density samples\,\cite{pellet2022spin}.
 
Let us concentrate on the ground $m_s$=\,$\pm1$ states. Only $m_s$=\,$\pm1$ states are taken into account since these are the states for which the transition energies to $m_s=0$ intersect/cross at zero-field. The ground state splitting for transition energies of different axes differs when a transverse field is present. As a result, there are multiple crossings between the transition energies at various values of $B_z$. The magnitude of the transverse field and the azimuth angle $\phi$ affect positions of crossings. When the field is scanned along [111] in the presence of 2\,mT of transverse field along $\hat{x}$ (this direction is defined so that one of the carbons associated with the NV center is in the x-z plane), the transition energies behave as shown in Fig.\,\ref{energy level}\,(b) for ground $\pm1$ states for all possible NV axes. There are fifteen transition-energy crossings, however some of them occur exactly at the same $B_z$ such that only five cross-relaxation features are expected. Note that for the calculations shown in Fig.\,\ref{energy level} (for 2\,mT of transverse field), we neglected the parameter $E$. While in general, up to six crossing positions can be observed when the [111] axis of the diamond is aligned with the z-axis of the setup.

\section{Experimental setup}
\begin{figure}
\includegraphics[width=3.0in]{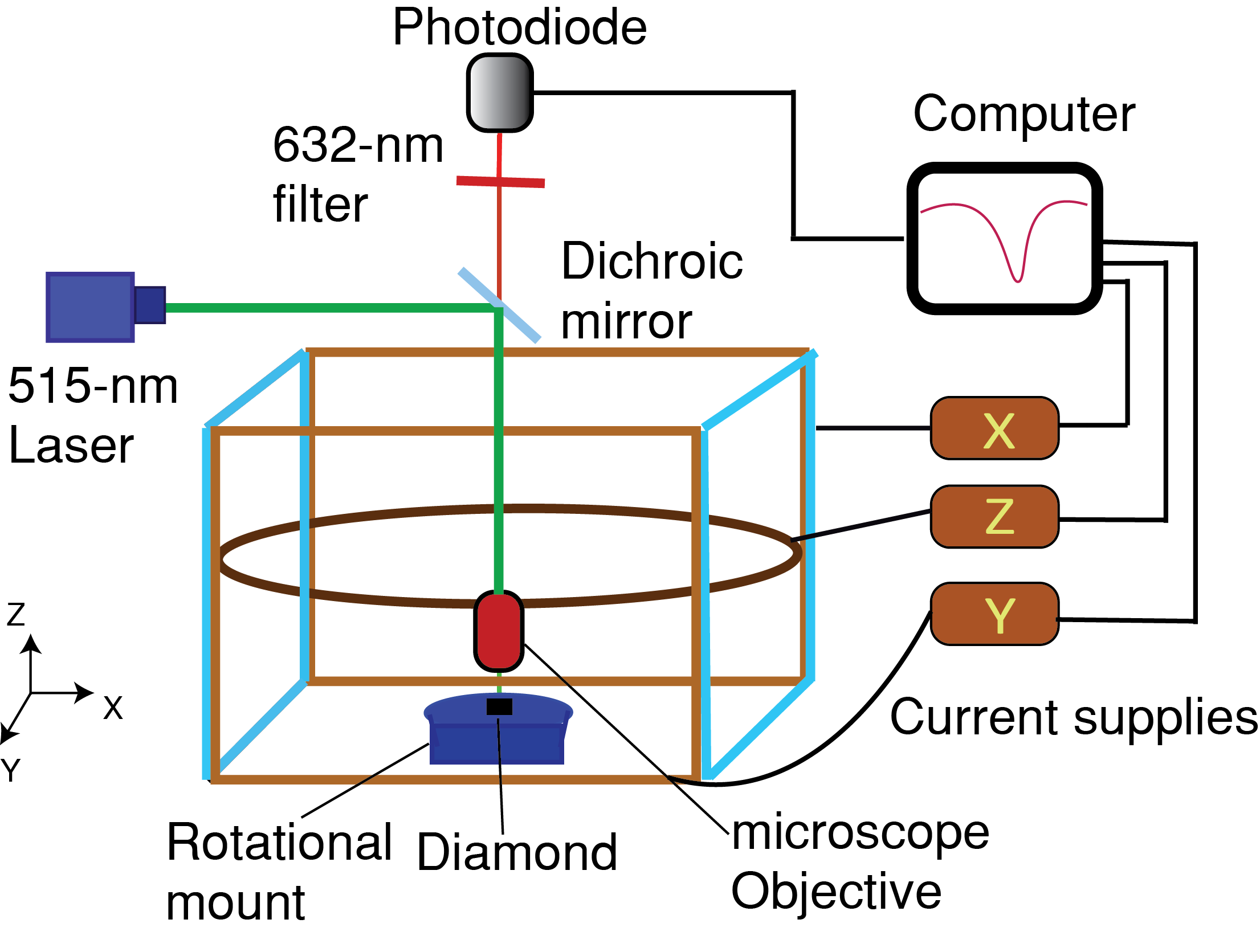}
\caption{Schematic diagram of experimental setup.
\label{schematic}
}

\end{figure}

\setstcolor{red}
To study the cross-relaxation features, we use a home-built wide-field fluorescence microscope (Fig.\,\ref{schematic}), where a laser (Toptica iBeam smart) with the output wavelength of 515\,nm is employed. The diamond is mounted on a rotational mount, and the laser beam is reflected via a dichroic mirror. The beam is then focused into the diamond using a microscope objective (Olympus PLN Plan Achromat 10x Microscope Objective, 0.25\,NA, 10.6\,mm WD). The NV-center fluorescence is collected using the same objective, passing the dichroic mirror and a long-pass filter to remove the green reflection of the laser light from diamond. The fluorescence is detected with a photo-diode (APD120A/M - Si Avalanche Photodetector, 400 - 1000 nm). Electromagnetic coils are used to apply a magnetic field in three different directions.

\begin{figure}
\includegraphics[width=3.0in]{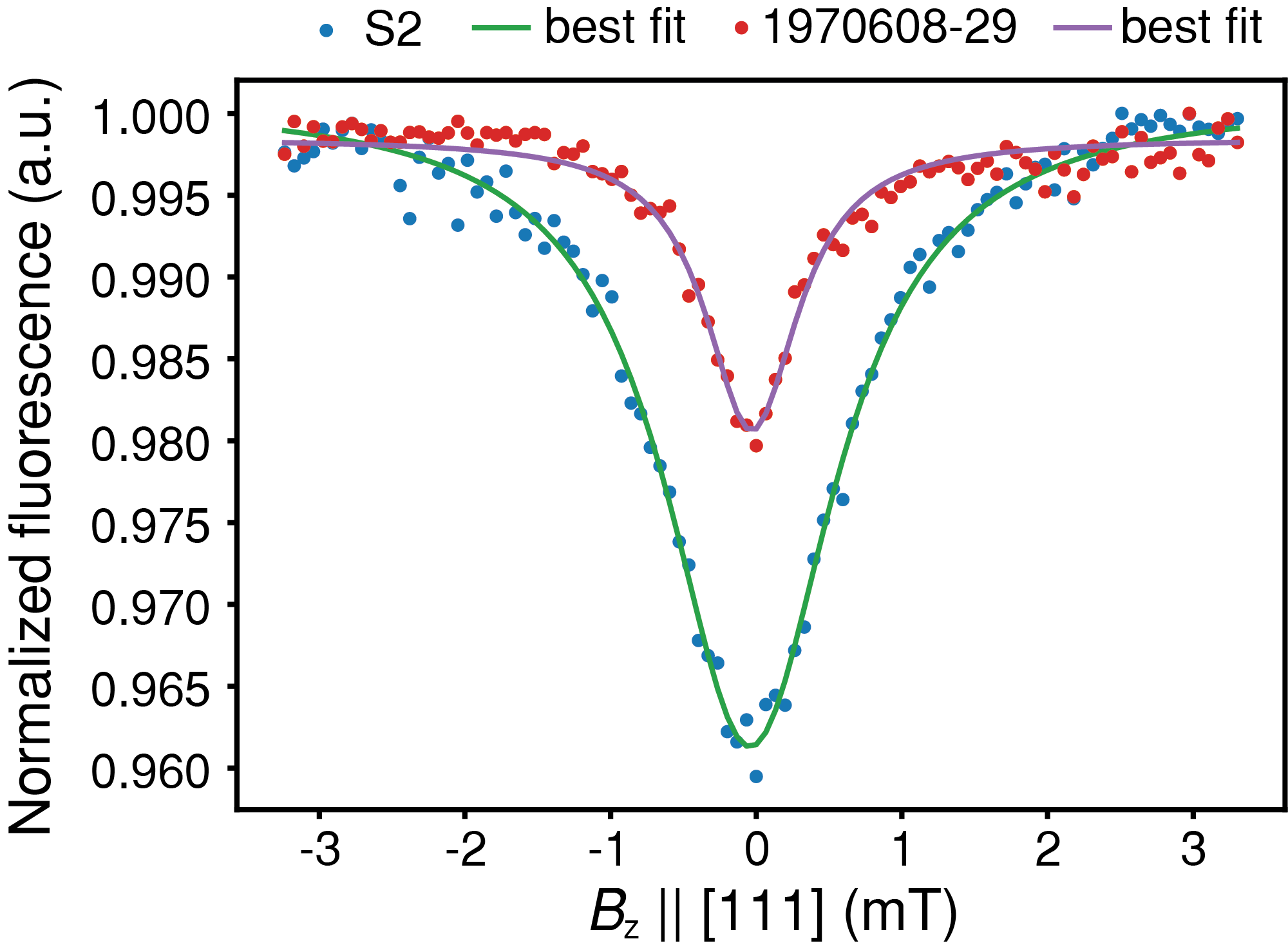}
\caption{Zero-fields features recorded on two different diamond samples while magnetic field along $\hat z$. For sample S2 (16\,ppm NV density), full width at half maximum (FWHM) and the contrast are 1.42(3)\,mT and 3.95(4)\,\%, respectively. For sample 1970608-29 (3.8\,ppm NV density), these are 0.78(4)\,mT and 1.79(5)\,\%.
\label{zero_field}
}
\end{figure}

\section{Results}

\subsection{Characterization of zero-field cross-relaxation feature}\label{characterization}

We explored the properties of the cross-relaxation feature at zero field in different diamond samples with varying NV density. During these experiments, neither microwave nor radio-frequency fields were applied. We only observed the cross-relaxation feature at zero-field in diamonds with NV density $\gtrsim0.9$\,ppm. Moreover, linewidth and contrast of the zero-field cross-relaxation feature depends on the NV-center density. In the examples in Fig.\,\ref{zero_field}, both linewidth and contrast increase with NV-center density. The contrast of the feature also depends on the laser intensity. Table\,\ref{bulk} summarizes the characteristics of the observed zero-field cross-relaxation feature in different samples. Characterization of the the zero-field cross-relaxation feature was performed with a laser spot diameter of 50\,$\mu$m.  For the bulk NVs, the optimum contrast is observed at different values of laser intensity for each sample. Higher laser intensity is used in the case of thin film NV samples because the signal-to-noise ratio for such samples is low at reduced power and the contrast increases with laser intensity. The NV density of the sumi\_300 kev sample is estimated based on a comparative NV fluorescence measurement with respect to the 2170612-13 sample. The linewidth and contrast are slightly position-dependent 
because of the variation of NV density over the sample. The lowest (i.e. most favorable) ratio of linewidth and contrast is observed for 3.7\,ppm-NV-concentration bulk sample\,(``George").

 \begin{figure*}
\includegraphics[width=7.0in]{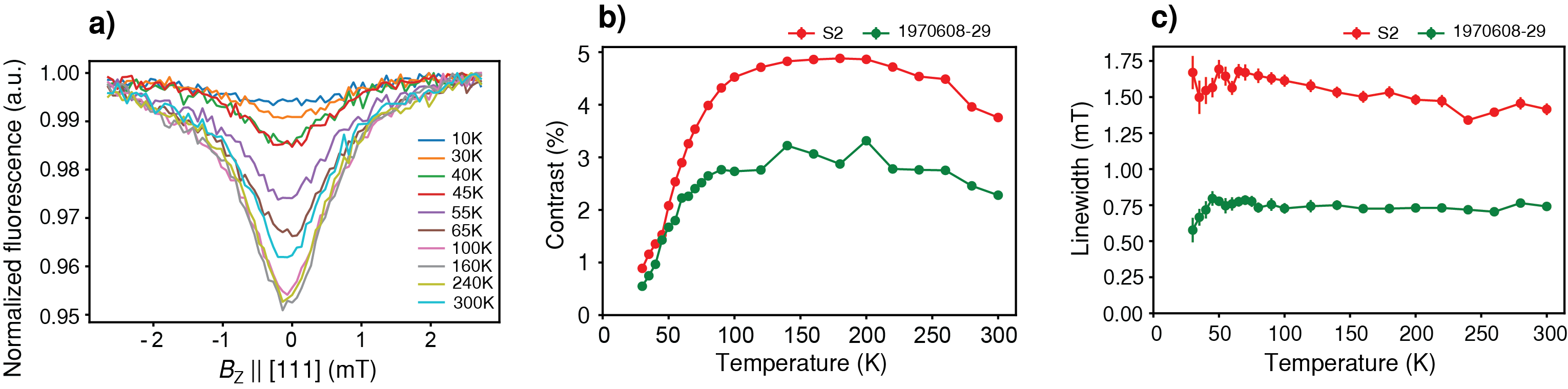}
\caption{Temperature dependence of the zero-field cross-relaxation feature. a) zero-field cross-relaxation feature at selected temperatures for sample S2. The zero-field cross-relaxation feature becomes observable at $\approx$\,5\,K. b), c) Linewidth and contrast dependence on temperature for sample S2 and 1970608-29. The features are Lorentzian fitted with 95\,{\%} confidence interval for each temperature to extract the linewidth and contrast.}\label{temp}
\end{figure*}
 
\subsection{Temperature dependence of zero-field cross-relaxation feature}\label{temperature}
We observed that the sample temperature affects the contrast of the zero-field cross-relaxation feature. The temperature dependence was investigated on two samples cut along (111) with the results shown in Fig.\,\ref{temp}. 

There are several effects that are of note. The zero-field signal is apparent at 5\,K which represents the lowest temperature achievable within our setup. However, this feature exhibits a significantly diminished contrast below $\approx$\,20\,K. The contrast increases with temperature and then declines above $\approx 250$\,K, retaining a sizable value at room temperature (Fig.\,\ref{temp} b). The highest contrast is observed in the range of 180-200\,K. The linewidth is roughly constant over the entire temperature range (Fig.\,\ref{temp} c). 
The high-temperature behavior is likely due to the increase of the longitudinal relaxation rate with temperature.

\begin{figure}[h!]
\includegraphics[width=3.4in]{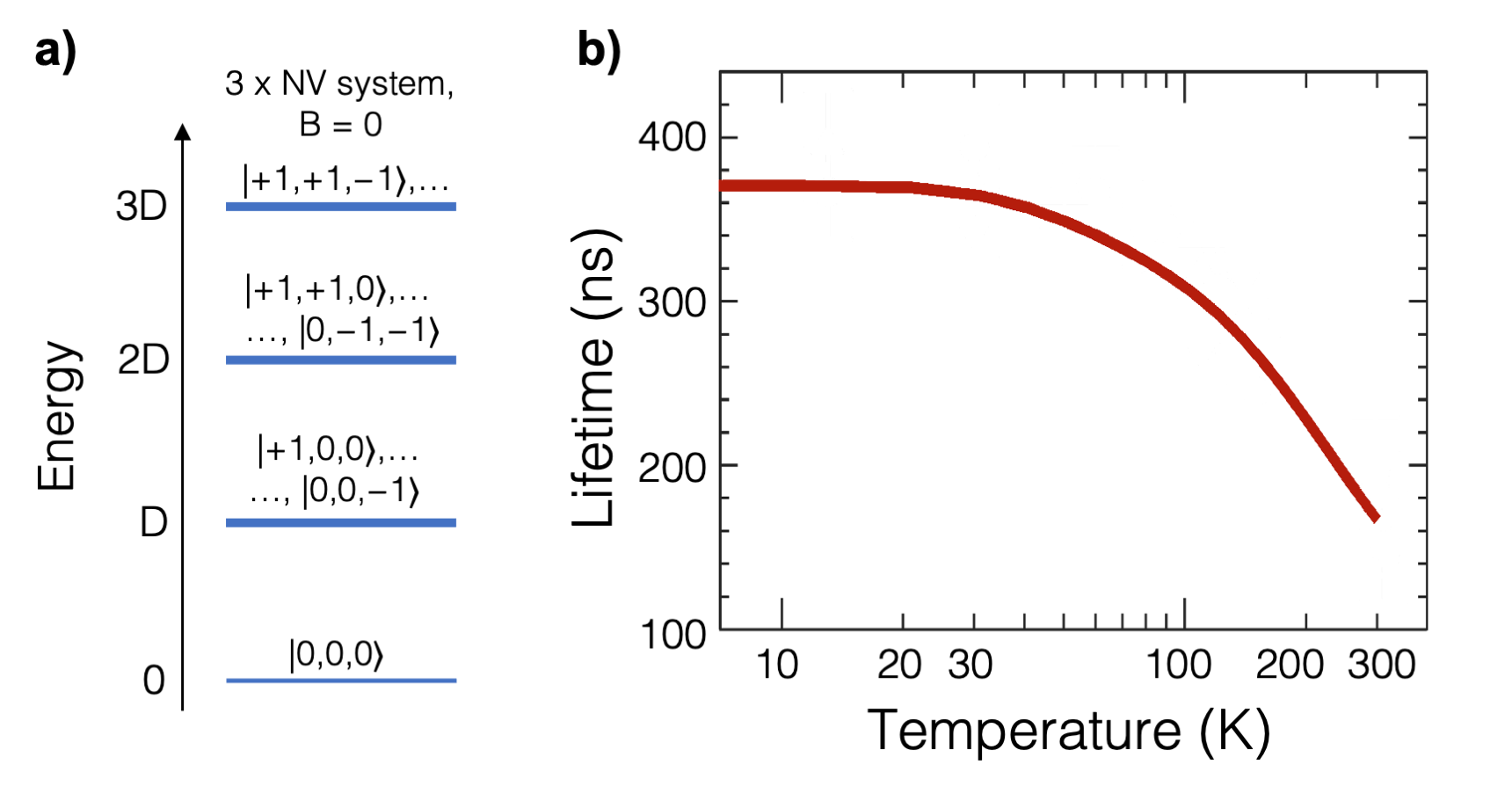}
\caption{a) Energy levels of three coupled NV centers and  b) temperature dependence of the lifetime of the $^{1}E$ NV shelving state\,\cite{thiering_theory_2018, PhysRevB.82.201202}.} \label{figth}
\end{figure}

In order to understand the observed 
temperature dependence of the contrast of the zero-field cross-relaxation feature at cryogenic temperatures, we need to identify the factors affecting the contrast. To do so we first study the energy-level structure of coupled NV centers, see Fig.\,\ref{figth}(a). When hyperfine, strain and electric-field-induced splittings of the spin states are neglected, the energy levels of a many-NV system fall into branches with a ladder-like structure, where the spacing of the steps is equal to the zero-field splitting (ZFS) parameter $D$. It is important to note the degeneracy of the branches. The lowest-lying $\left| 0,0,…,0  \right\rangle $ state is non-degenerate, even for NV centers of different orientations. The higher-lying states are, however, degenerate and include mixed $\left| 0 \right \rangle$ and $\left| \pm 1 \right \rangle$ states, see Fig.\,\ref{figth}(a) illustrating this for three NV centers. When all NV centers are completely polarized to $\ket{0}$, the many-spin system populates the non-degenerate (lowest energy) state. 
Spin-relaxation effects may connect the lower-lying state with-higher lying states; however, such processes are suppressed by the large value of the ZFS compared to the typical strength of dipole-dipole interactions. 

Drop of the NV polarization, i.e., population of the higher-lying energy states, enables spin flip-flop among the NV enters. However, the averaged probability of finding the NV centers in $\left| 0 \right \rangle$ does not change even in this case. Dipolar spin relaxation alone cannot account for the zero-field cross-relaxation feature.

To understand depolarization in a dense NV ensemble, we need to utilize the concept of spin-fluctuators as explained in Ref.\,\cite{PhysRevLett.118.093601}. Spin-fluctuators are NV centers with a short lived ground-state electron spin, through which a polarized, dipolar coupled NV bath can dissipate its energy faster than through the conventional spin-lattice relaxation. At non-zero magnetic field, only NV centers and fluctuators along the same axes couple. At zero magnetic field, however, all NV centers couple to each other, which increases the effective density of the NV centers and the spin-fluctuator bath. In turn, this gives rise to the zero-field cross-relaxation feature. The contrast of the zero-field cross-relaxation feature thus depends on the density of the NV centers as well as the ratio of the short lived spin fluctuators and the ``normal” long-lived NV centers. Since the former is not temperature but sample dependent, the latter as a possible temperature-dependent factor.

As discussed in Ref.\,\onlinecite{PhysRevLett.118.093601}, spin-fluctuators can be close NV centers of different charge states, where electron (charge) tunneling between the centers can shorten the spin-state lifetime in the ground state of the negatively charged NV center. Therefore, the fluctuator per normal NV center ratio and, correspondingly, the contrast of the zero-field cross-relaxation feature depend on the NV($-$)/NV($0$) ratio\,\cite{supp}, which is temperature, excitation-power, and sample dependent. Considering the applied 515\,nm excitation, which is energetic enough to ionize the NV center from the $^1E$ shelving state (see, for example, Ref.\,\onlinecite{Bockstedte2018}), we attribute the temperature dependence of the contrast to the temperature dependence of the lifetime of the $^1E$ electronic state of the NV center. The $^1E$ electronic state couples to the higher-lying $^1A_1$  state through the pseudo Jahn-Teller effect and to the even higher lying $^1E^{\prime}$ state through the dynamics Jahn-Teller effect\,\cite{thiering_theory_2018}. The first excited vibronic level can be found 16\,meV above the lowest vibronic level of $^1E$ state, which gives rise to the characteristic temperature dependence, depicted for example for the lifetime of $^1E$ in Fig.\,\ref{figth}\,(b).  Boltzmann occupation of the excited vibronic levels significantly increases beyond 20\,K, which on one hand shortens the lifetime of the $^1E$ state, see Fig.\,\ref{figth}\,(b), and on the other hand increases the photo-ionization rate from $^1E$ state towards the conduction band.  We attribute the onset of the zero-field contrast at around 20\,K to the increasing ionization rate due to the thermal occupation of the vibronic excited energy levels of the $^1E$ state (e.g., such an effect is discussed for silicon carbide divacancy centers in Ref.\,\onlinecite{Csore2022}), whereas the shortening of the $^1E$ lifetime, i.e., intersystem crossing towards the ground state of NV($-$), will compete with this photo-ionization process at elevated temperatures. Quantitative simulation of the contrast requires additional information on the dynamics of the spin fluctuators, which is the subject of further investigations.

From the study of temperature dependence and the above discussion we conclude that the NV($-$)/NV($0$) ratio governs the temperature dependence of the zero-field cross-relaxation feature contrast. Sensitive dependence of this parameter on excitation power and wavelength gives pathways to engineer the contrast of zero-field cross-relaxation feature without the need for microwave excitations.

 \begin{figure*}
\includegraphics[width=7.0in]{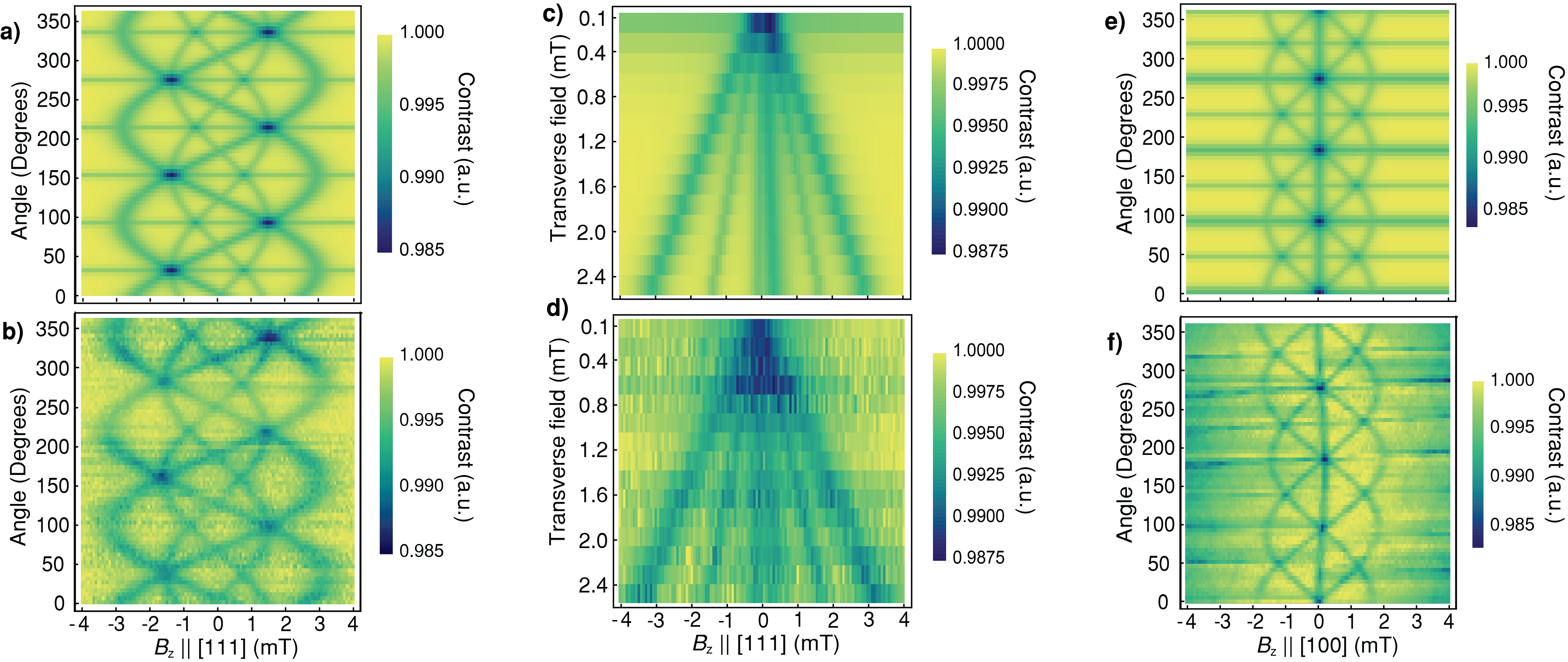}
\caption{Simulation\,(upper row) and experimental measurement\,(lower row) of the cross-relaxation features in the presence of a transverse field of 2\,mT. (a) Simulation, (b) measurement of the transition energy difference between $m_s$= +1 and -1 states for all four NV axes. Here $B_z$ is along [111] and the azimuthal angle of the transverse field is $\phi$. 
(c), (d) Varying the transverse field and $B_z$ at an angle of 5$^\circ$.  
(e), (f) Transition energy difference between $m_s$= +1 and -1 states when $B_z$ is along [100] at varying angles when 16\,mT of transverse field is applied.} 

\label{111simulation}
\end{figure*}

\subsection{NV-NV cross-relaxation in the presence of a transverse field}\label{transverse_field}

This section describes calculations and experimentally measured additional cross-relaxation features observed when $B_z$ is along [111] (and [100] later on) when applying 2\,mT (1.6\,mT for [100]) transverse field.

 Figure\,\ref{111simulation}(a) shows a density plot of the simulated contrast based on the numerically evaluated transition frequency differences as a function of $B_z$ along [111] and the azimuthal angle of the transverse magnetic field (2\,mT). Here, we assume that the cross-relaxation features have a Lorentzian line shape. The parameters of the Lorentzian function are adjusted to reflect the measured data.
There are multiple crossings between the transition energies for differently oriented NVs at different values of $B_z$. The position of the crossings and the overlap between the transition energies depend on $\phi$ when the diamond is rotated around $\hat{z}$ with a transverse field present. When $B_z$ is along [111]\,(Fig.\,\ref{111simulation}\,(a)), for every multiple of 60$^\circ$, starting at 30$^\circ$, transitions for NVs along two of three non-aligned crystallographic axes overlap each another. These are the orientations where the crossings between the transition energies at the three values of $B_z$ occur. Also, two out of three inequivalent transition energies successively cross each other at $B_z$\,=\,0 for every multiple of 60$^\circ$, starting from 0$^\circ$. For each of these values of $\phi$, crossings occur at five distinct values of $B_z$, Fig.\,\ref{energy level}\,(b) shows one of those cases when $\phi$\,= 0$^\circ$.

Figure\,\ref{111simulation} (b) shows experimentally measured cross-relaxation patterns between NVs oriented along different axes. 
For every multiple of 60$^\circ$, starting from 30$^\circ$ there is an overlap of three separate cross-relaxation features at -1.4\,mT and +1.4\,mT and of two separate features at +\,0.7\,mT and -\,0.7\,mT. As a result, cross-relaxation features are observed at only three values of $B_z$ at these angles. For multiples of 60$^\circ$ starting from 0$^\circ$, there is an overlap of two separate features at $B_z$\,=\,0 so that five crossings are observed at these angles. There are always six cross-relaxation values of $B_z$ for the rest of the angles but these values depend on the angle $\phi$. The observed cross-relaxation features when we apply transverse field have narrower linewidth than the zero transverse-field cross-relaxation feature, which can be attributed to the effect of electric field, strain, and additional cross-relaxation channels in the sample. We also note the different widths of the lines in the figure. This is related to the difference in the derivatives of the crossing transition energies as a function of $B_z$. Note also that one can estimate the direction of the magnetic field from the positions of the crossings, which is useful for vector magnetometry applications\,\cite{PhysRevA.100.043844}.

 When $B_z$ is scanned for each value of the  transverse field, the splitting of the cross-relaxation features linearly depends upon the transverse field i.e. all of these cross-relaxation features are magnetically sensitive. (An exception are crossings occurring at $B_z$\,=\,0 for certain values of the azimuthal angle.) 
Figure\,\ref{111simulation}(c), (d) shows the simulated and experimentally measured cross-relaxation features for $\phi\approx$\,5$^\circ$ angle. It shows the linear dependence on the transverse magnetic field. A limitation of these extra cross-relaxation features (other than the cross-relaxation feature at zero longitudinal field) for magnetometry applications requiring operation near-zero field is that application of a 0.5-1.0\,mT transverse field is necessary to clearly resolve these features.

\begin{figure}[h]

\includegraphics[width=3.30in]{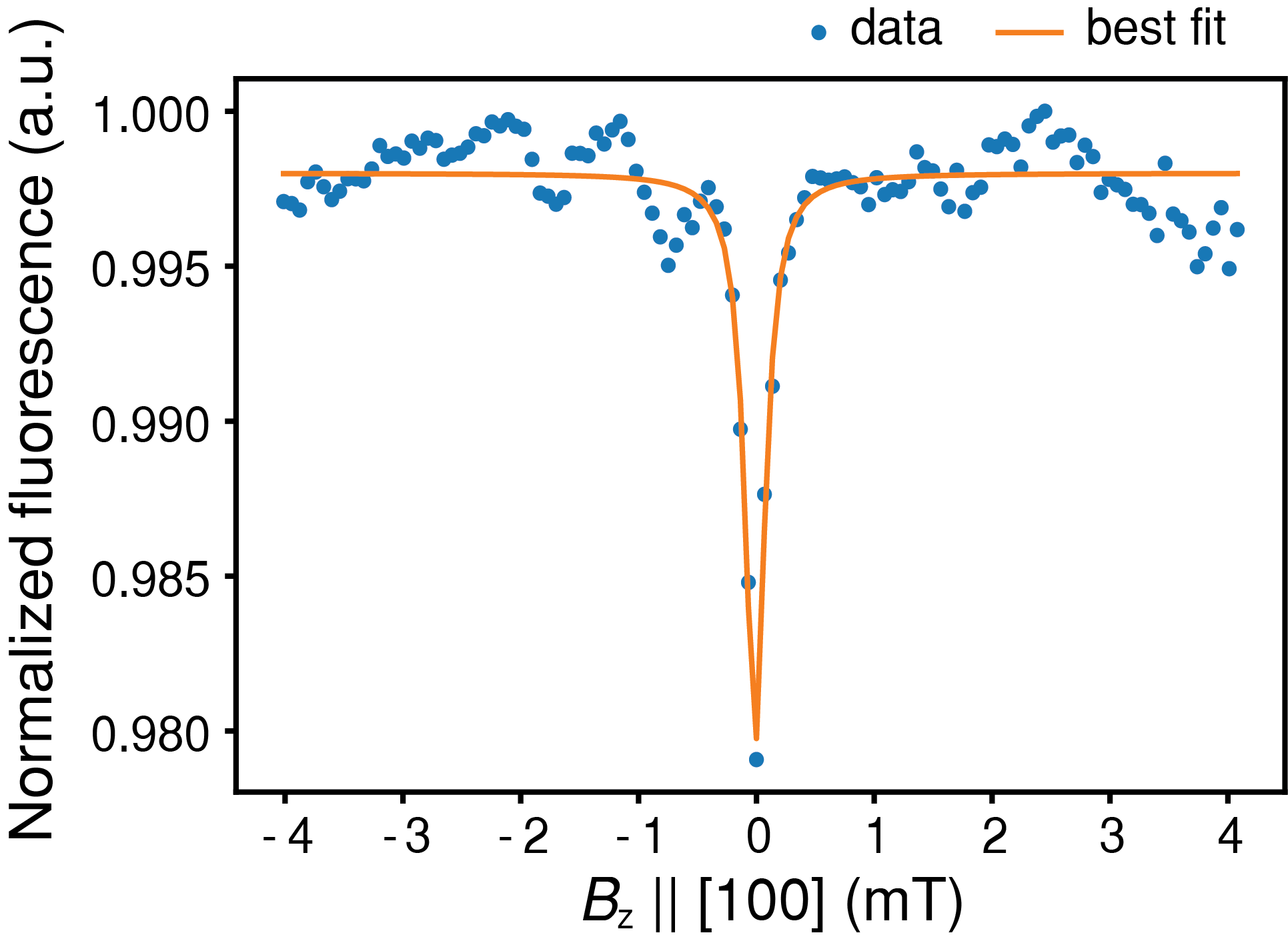}
\caption{Dominant central cross-relaxation feature observed while applying 1.6\,mT of transverse field at an azimuthal angle of $\approx$\,90$^\circ$. The FWHM and the contrast for this feature are 0.20(2)\,mT and 1.85(10)\%, respectively. 
\label{90deg}
}
\end{figure}

Similarly, in the case of $B_z$ along [100]\,(Fig.\,\ref{111simulation}\,(e)), due to the crystal symmetry, transverse fields are equal for two groups of NVs that have pairs of transition energies crossing each other at $B_z$\,=\,0 for every multiple of 90$^\circ$, starting at 0$^\circ$. Since, these two pairs of transition energies intersect at $B_z$\,=\,0 there is just one crossing between the transition energies at $B_z$\,=\,0. 
Figure\,\ref{90deg} shows the dominant single cross-relaxation feature at one of such angles, $\approx$\,90$^\circ$. At this angle, the contrast is larger than that of the zero transverse-field cross-relaxation feature of the same sample\,(``George"). Other small features are still observed, likely because of a small misalignment with respect to the transverse field. Additionally, two out of every four transition energies successively overlap at every multiple of 45$^\circ$, resulting in crossings at three different values of $B_z$.
Figure\,\ref{111simulation}(e) shows the density plot of transition-frequency differences as a function of $B_z$ along [100] and the azimuthal angle of the transverse magnetic field at 1.6\,mT. Figure\,\ref{111simulation}\,(f) shows the experimentally measured cross-relaxations features. Since the transition energies for each angle cross at $B_z$\,=\,0 a cross-relaxation feature is always visible at zero-field. There are five cross-relaxation positions for the remaining angles. In our setup, while rotating the diamond, there is a slight translation. Therefore, if we compare the experimentally measured data in Fig.\,\ref{111simulation}(f) with the calculations in Fig.\,\ref{111simulation}(e), we see a slight displacement of the features which, however, does not obscure the overall good agreement.

The cross-relaxation features in the presence of a transverse field are measured  on diamond sample ``S2" for [111]  and on sample ``George"  for [100] (see Tab.\,\ref{bulk}).
\setstcolor{red}

The photon shot noise limited sensitivity is approximated as $\eta \approx \sqrt{2}\Delta B/C/\sqrt{I_0}$, where $\Delta B$ = 1.2\,mT is the linewidth and $C$ = 3.0\% is the contrast for the zero-field cross-relaxation feature of the S2 sample. The photon detection rate $I_0$\,= 2.4\,$\times\,10^{13}$\,/s. For these values, the estimated photon shot noise limited sensitivity is around 2.7\,\,nT/$\sqrt\textrm{Hz}$ while the measured sensitivity, as detailed in the supplementary materials\,\cite{supp}, is approximately 110\,nT/$\sqrt\textrm{Hz}$, exhibiting a relatively consistent noise floor across frequencies. This observation suggests that the performance is limited not by shot noise, but rather by other sources of noise such as laser noise, ambient magnetic noise and etc. Enhancements in sensitivity can be attained through the optimization of light power and the augmentation of light-collection efficiency \cite{Omar:23, barry2023sensitive}. This sensitivity is sufficient for applications such as studying superconducting vortices and magnetic properties of magnetic materials\,\cite{PhysRevApplied.10.034032, PhysRevApplied.15.024040}.

\subsection{Conclusion}\label{conclusion}
\setstcolor{red}

We have investigated the zero-field cross-relaxation feature with respect to the NV density and the axis of sample cut. The NV density influences both the linewidth and the contrast of this feature. The temperature dependence of the zero-field cross-relaxation feature was studied. The feature is more prominent above 20\,K. The contrast increases with temperature, and gradually decreases from about 250\,K\ref{temp} . In the presence of a transverse field, various cross-relaxation features are detected due to the dipolar interaction between differently oriented NVs providing depolarization channels. The number and location of cross-relaxation features are determined by the orientation of the magnetic field with respect to the crystal axes, azimuthal angle and the strength of the transverse field. These cross-relaxation features follow specific pattern (well reproduced by our theoretical calculations) when the azimuthal angle is changed with respect to the transverse field. These features have narrower linewidth than the zero-field cross-relaxation feature. Moreover, higher contrast is observed at certain angles where there is only one decay channel available to depolarize.

These results will be used in scalar- and vector-magnetometry applications, forming the basis of a practical near-zero-field microwave-free or even all-optical technique.

\subsection{Acknowledgement} 
We thank Junichi Isoya for helpful discussions and providing a sample. This work was supported by the European Commission’s Horizon Europe Framework Program under the Research and Innovation Action MUQUABIS GA no. 101070546, by the German Federal Ministry of Education and Research (BMBF) within the Quantumtechnologien program (Grant No.\,13N15064 and Grant No.\,13N16455), by the German DFG, Project SFB 1552 "Defekte und Defektkontrolle in weicher Materie”, funding by the Carls Zeiss-Stiftung (HYMMS P2022-03-044) and by the DAAD/JSPS 2021-23 cooperation grant No.\,57569949. A.G.\ and V.I.\ acknowledge support from the Ministry of Culture and Innovation and the National Research, Development and Innovation Office within the Quantum Information National Laboratory of Hungary (Grant No.\ 2022-2.1.1-NL-2022-00004). V.I. acknowledges support from the National Research, Development, and Innovation Office of Hungary (NKFIH) (Grant No.\ FK 145395) and the Knut and Alice Wallenberg Foundation through WBSQD2 project (Grant No.\ 2018.0071). A.G.\ acknowledges the Hungarian NKFIH grant No.\ KKP129866 of the National Excellence Program of Quantum-coherent materials project, the EU HE EIC Pathfinder project QuMicro (Grant No.\ 101046911), and the QuantERA II project MAESTRO.

\bibliography{bibliography.bib}

\end{document}